\documentclass{article}
\usepackage{spconf}
\usepackage{amsmath,amssymb,amsfonts}
\usepackage{algorithmic}
\usepackage{graphicx}
\usepackage[ruled]{algorithm2e}
\usepackage{booktabs}
\usepackage{textcomp}
\usepackage{hyperref}

\usepackage{enumitem}
\setlist{nosep, leftmargin=14pt}

\title{ISLE: An Intelligent Streaming framework for High-Throughput AI Inference in Medical Imaging}
\name{Pranav Kulkarni, Sean Garin, Adway Kanhere, Eliot Siegel, Paul H. Yi, Vishwa S. Parekh\thanks{Corresponding author: vparekh@som.umaryland.edu}}
\address{University of Maryland Medical Intelligent Imaging (UM2ii) Center,\\University of Maryland School of Medicine, Baltimore, MD 21201}
\begin{document}
%
\maketitle
\begin{abstract}
As the adoption of Artificial Intelligence (AI) systems within the clinical environment grows, limitations in bandwidth and compute can create communication bottlenecks when streaming imaging data, leading to delays in patient care and increased cost. As such, healthcare providers and AI vendors will require greater computational infrastructure, therefore dramatically increasing costs. To that end, we developed ISLE, an intelligent streaming framework for high-throughput, compute- and bandwidth- optimized, and cost effective AI inference for clinical decision making at scale. In our experiments, ISLE on average reduced data transmission by 98.02\% and decoding time by 98.09\%, while increasing throughput by 2,730\%. We show that ISLE results in faster turnaround times, and reduced overall cost of data, transmission, and compute, without negatively impacting clinical decision making using AI systems.
\end{abstract}
\section{Introduction}

Artificial Intelligence (AI) systems are increasingly being developed and deployed in radiology across different applications involving different imaging modalities and body parts. Deploying AI solutions in radiology typically involves healthcare providers streaming large volumes of medical image data to AI vendors for analysis. However, limitations in bandwidth can potentially create communication bottlenecks when streaming medical images over the internet, leading to delays in patient diagnosis and treatment. To address these communication bottlenecks, healthcare providers generally use lossless compression codecs  (e.g., JPEG-LS, JPEG-2000) to compress DICOM imaging data before sending them over the network. 

While lossless compression codecs are able to reduce the size of imaging data for transmission, they still require healthcare providers to send full resolution imaging data to AI vendors for analysis. However, a majority of AI models, especially deep learning classification models, are developed with significantly lower resolution images (e.g., 224x224 for transfer learning) compared to standard acquisition resolutions. Therefore, if we could intelligently stream medical imaging data only at the necessary resolution for AI inference, it could significantly reduce the computational infrastructure required for clinical deployment of these AI systems, democratizing their deployment across both small and large healthcare systems while also reducing the overall carbon footprint of clinical AI deployments. 

\begin{figure}[!t]
    \centering
    \includegraphics[width=0.9\linewidth]{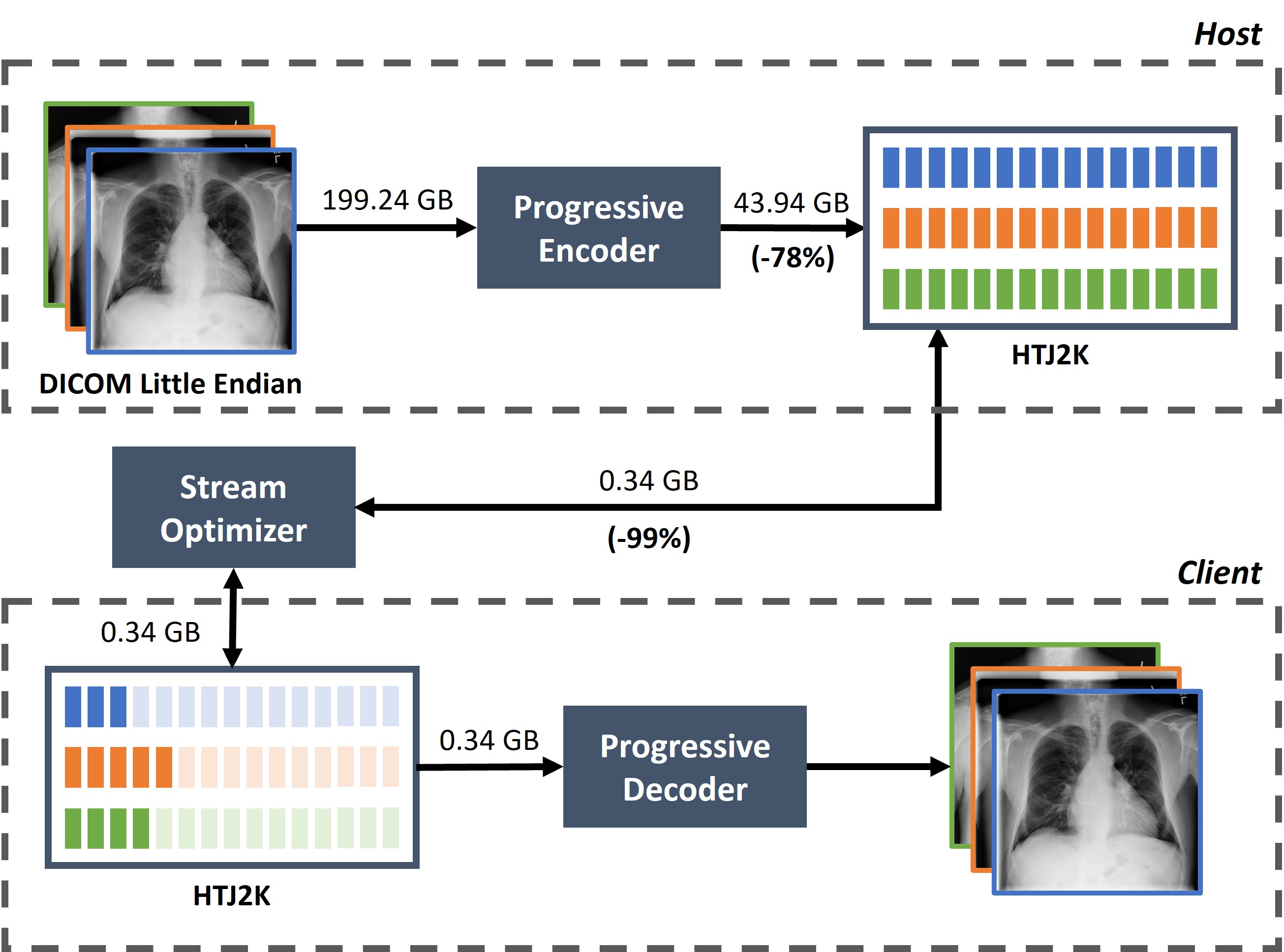}
    \caption{An overview of ISLE, our proposed intelligent streaming framework for high-throughput AI inference for clinical decision making at scale. }
    \label{fig:overview}
\end{figure}

\begin{figure}[!t]
    \centering
    \includegraphics[width=0.9\linewidth]{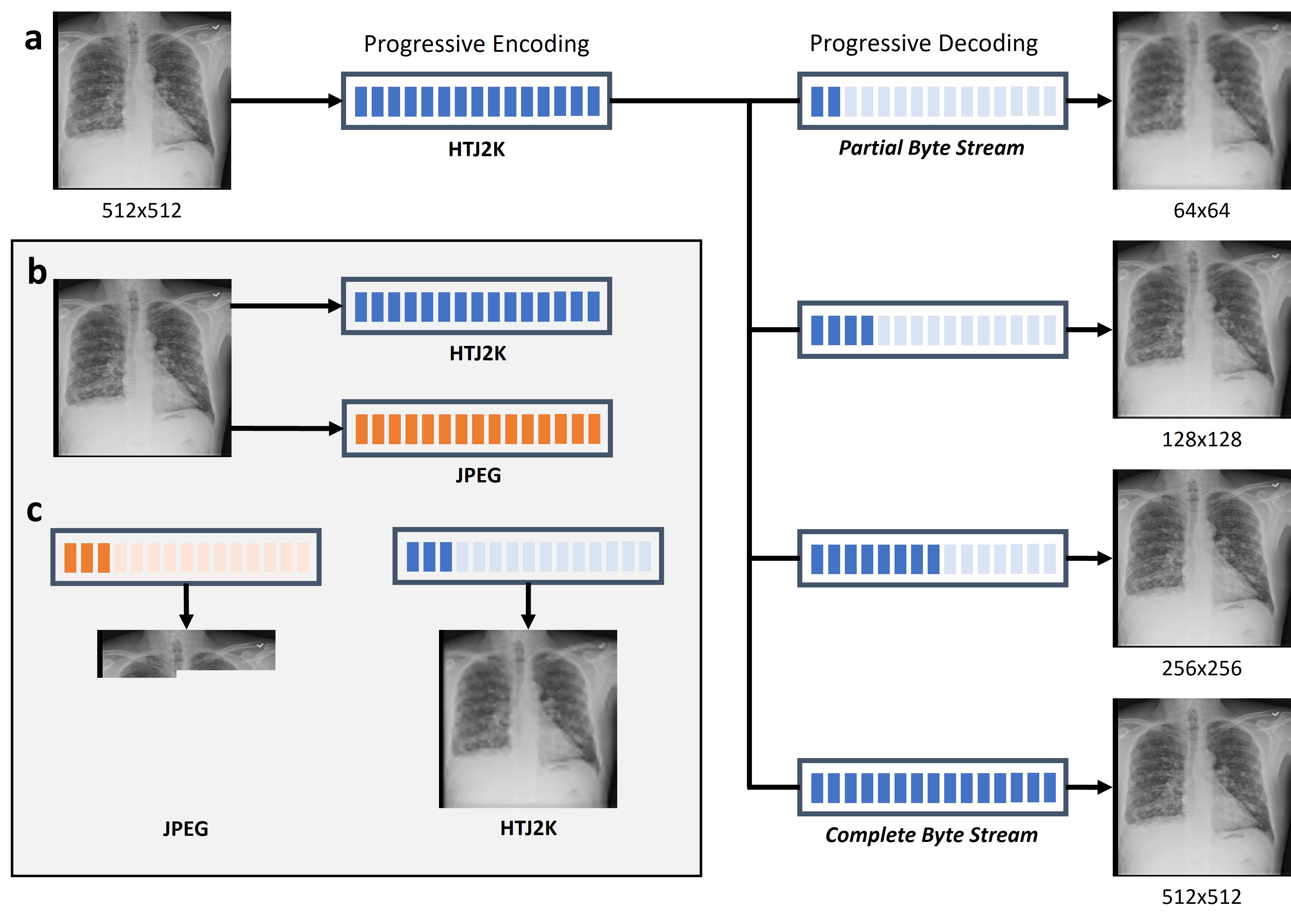}
    \caption{\textbf{(a)} Illustration of the progressive encoding of a chest x-ray and subsequent progressive decoding at different decompositions (i.e., sub-resolutions) by selecting different subsets of the image byte stream. \textbf{(b)} Suppose the same chest x-ray is encoded using JPEG (sequential) and HTJ2K (progressive) encoders, followed by decoding of partial byte streams from both formats. \textbf{(a)} The JPEG image results in an incomplete image whereas HTJ2K image is visible, albeit lower in resolution.}
    \label{fig:progressive}
\end{figure}

To that end, we developed ISLE, an intelligent streaming framework for high-throughput, compute- and bandwidth- optimized, and cost effective AI inference for clinical decision making at scale (Fig. \ref{fig:overview}). The ISLE framework is based on progressive encoding coupled with intelligent streaming, a technology that has revolutionized the way we consume media content over the internet. Progressive encoding works by encoding the imaging data into a series of decompositions, each with an increasing level of detail (Fig. \ref{fig:progressive}(a)). However, the computational overhead of progressive encoding has held back its adoption in medical imaging beyond niche web-based applications \cite{foos2000jpeg}. 

Recently, the growing popularity of High-Throughput JPEG 2000, a state-of-the-art progressive codec, has renewed interest in progressive resolution with its integration into the DICOM transfer syntax and large-scale medical image cloud providers such as Amazon Web Services' HealthImaging \cite{hafeyhtj2k,htj2k_dicom,amazonhealthlake}. Our framework is built on top of HTJ2K to determine and intelligently stream an optimal subset of the image byte stream that can deliver the same performance as the full image byte stream, while enabling faster turnaround times with reduced overall cost of data storage and transmission. In this work, we evaluate the ISLE framework for optimal streaming of 2D chest x-ray (CXR) datasets for diagnostic classification, comparing it against conventional and progressive streaming frameworks.

\section{Methods}

\subsection{High-Throughput JPEG 2000 (HTJ2K)}

High-Throughput JPEG 2000 (HTJ2K) is an extension to the existing JPEG 2000 codec, that -- unlike the original JPEG standard -- enables faster decoding, higher quality, lossless encoding with high scalability, and progressive decoding of images \cite{foos2000jpeg,taubman2019high}. Imaging data can be progressively encoded using HTJ2K as $N$ decompositions (i.e., sub-resolutions), each with a native resolution as a factor of $2^N$ of the full image resolution. Unlike the sequential streams of the original JPEG standard, HTJ2K streams are transmitted as a sequence of these $N$ decompositions (Fig. \ref{fig:progressive}(a)). For example, a 2D CXR with a resolution of 256x256 can be progressively encoded as 3 decompositions with resolutions 64x64, 128x128, and 256x256. This allows HTJ2K to not just achieve a high compression ratio, but also enable quick access to smaller resolution versions of the image, thus minimizing data duplication.

\subsection{Intelligent Streaming}
The ISLE framework consists of three primary components, as visualized in Figure \ref{fig:overview}. 

\subsubsection{Progressive Encoder}

The progressive encoder encodes imaging data into HTJ2K byte streams. We utilize OpenJPHpy, a Python wrapper for OpenJPH, an open-source implementation of the HTJ2K codec, to encode each image as lossless with 64x64 block size, $N$ decompositions (see Eq. \ref{eq:decomp}), and tile-part markers to identify the location of each decomposition within the complete HTJ2K byte stream \cite{openjphpy,openjph}.
\begin{equation} \label{eq:decomp}
N = \left\lfloor \log_{2} \left(\frac{\min (X,Y)}{\alpha}\right) \right\rfloor
\end{equation}
where $N$ is the number of decompositions to encode, $(X,Y)$ is the resolution of the image, and $\alpha$ is lower bound for the resolution of the first decomposition (i.e., the smallest sub-resolution). We arbitrarily set $\alpha=32$.

\subsubsection{Stream Optimizer}

The stream optimizer determines the optimal decomposition of an image that can be streamed without negatively impacting the performance of an AI model. More specifically, only the subset of the image byte stream sufficient to reconstruct the image at the optimal decomposition is streamed to a client. In stark contrast to conventional imaging databases that transmit the entire image, streaming partial byte streams enables ISLE to transmit only the data required for the AI system, minimizing unnecessary data transmission.

The optimal decomposition is determined on the client-side based on two factors: 1) The architecture of the AI system (e.g., input image size, deep learning architecture, etc.), 2) The smallest decomposition for which there is no statistically significant difference in the AI system's performance using an internal validation set.

\subsubsection{Progressive Decoder}

The progressive decoder decodes the partial byte streams transmitted via the stream optimizer to reconstruct the optimal decomposition, with its corresponding resolution. While the HTJ2K specification does not explicitly support the decoding of partial byte streams, this unintended capability arises from the format's structure. On the other hand, conventional standards (e.g., JPEG, PNG, etc.) sequentially encode an image from the top-left to the bottom-right pixels, thus preventing images from being decoded from a partial byte stream in the same way that HTJ2K allows (Fig. \ref{fig:progressive}(b-c)).

\subsection{Datasets}

\subsubsection{NIH Chest X-Ray 14}

The NIH Chest X-Ray 14 dataset consists of $n=112,120$ frontal CXRs from 30,805 patients, encompassing a total size of 42 GB \cite{wang2017chestx}. All images in the dataset have a fixed resolution of 1024x1024, pixel intensities normalized to [0,255], and encoded in PNG (lossless) format. For deep learning, the dataset was randomly divided into training (70\%, $n=78,075$), validation (10\%, $n=11,079$), and testing (20\%, $n=22,966$) splits with no patient leakage.

\subsubsection{CheXpert}

The CheXpert dataset consists of with $n=224,316$ frontal and lateral CXRs from 65,240 patients, encompassing a total size of 440 GB \cite{irvin2019chexpert}. Images in the dataset have a mean resolution of 2828x2320. All images have pixel intensities normalized to [0,255] and encoded in JPEG (lossy) format. For inference, we sampled $n=15,000$ frontal CXRs and all uncertain labels were treated as negatives.

\subsubsection{MIMIC-CXR-JPG}

The MIMIC-CXR-JPG dataset consists of $n=377,110$ frontal and lateral CXRs from 65,379 patients, encompassing a total size of 558 GB \cite{goldberger2000physiobank,johnson2019mimic}. Images in the dataset have a mean resolution of 2500×3056. All images have pixel intensities normalized to [0,255] and encoded in JPEG (lossy) format. For inference, we sampled $n=15,000$ frontal CXRs and all uncertain labels were treated as negatives.

Furthermore, the MIMIC-CXR dataset provides DICOM versions of the data, with imaging data encoded in little endian (uncompressed) format and encompassing a total size of 4.6 TB \cite{johnson2019mimicdicom}. For inference, the same $n=15,000$ CXRs were used and same preprocessing steps applied to the images. 

\subsection{Experiments}

We evaluate the ISLE framework for AI inference using 2D CXR classification model to detect abnormalities in the thorax. The model was trained using an internal dataset from a single site (i.e., NIH), the optimal decomposition was determined and validated using held-out internal test set from the same site (i.e., NIH), and then evaluated on unseen data transmitted by two different sites (i.e., CheXpert and MIMIC). The data transmitted, decode time, and throughput (number of images processed by the AI system per second) were measured as data metrics and compared to the non-ISLE streamed versions of the data. Decode time was measured using OpenCV (version 4.7) for JPEG and PNG formats, pydicom (version 2.3.1) for DICOM imaging data, and OpenJPHpy (version 0.1) for HTJ2K format.

We trained a multi-label DenseNet121 model, pre-trained with ImageNet, using transfer learning for 100 epochs with binary cross-entropy loss, batch size of 64, and learning rate of 5e-5. All CXRs were downsampled to 224x224 while maintaining aspect ratio (i.e., zero-padded), normalized between 0 and 1, and scaled to ImageNet statistics. Random augmentations (rotation, flip, zoom, and contrast) were applied to input images during training. All models were trained and evaluated using TensorFlow (version 2.8.1) and CUDA (version 12.0) on a NVIDIA RTX A6000 GPU.

Model performance was measured using the mean area under the receiver operating characteristic curve (AUROC) on the original dataset and across each HTJ2K decomposition (which includes ISLE's optimal decomposition). Paired one-tailed t-tests between the original data and each HTJ2K decomposition, with alternative hypothesis defined by the mean AUROC of the decomposition is less than the mean AUROC of the original data. Statistical significance was defined as $p<0.05$. In all cases, normality assumption was satisfied by $p>0.05$ using the Shapiro-Wilk test. 

\section{Results}

\begin{figure*}[!t]
    \centering
    \includegraphics[width=0.9\linewidth]{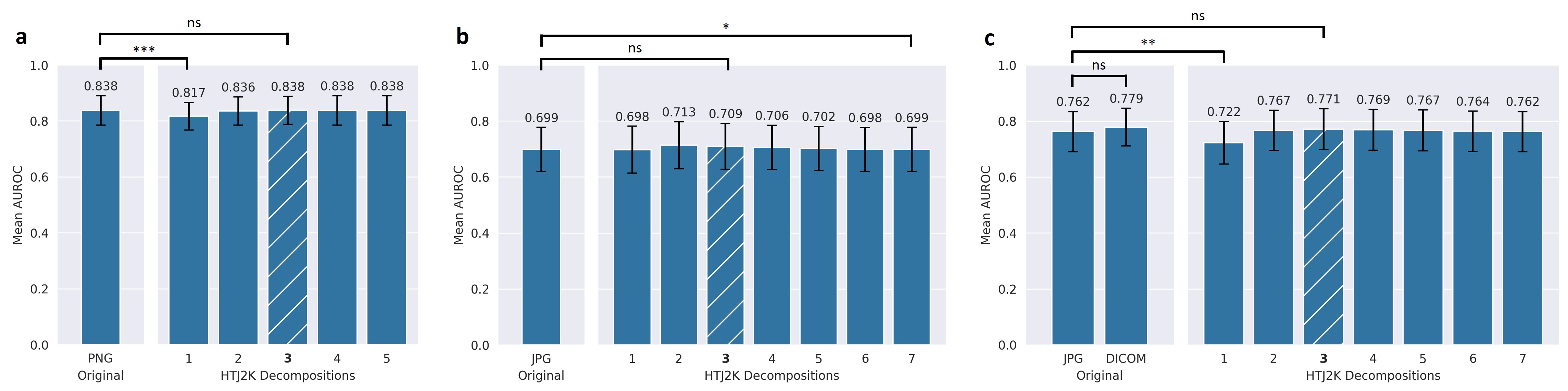}
    \caption{Mean AUROC scores on the original dataset and across each HTJ2K decomposition, including ISLE's optimal decomposition (hatched) on the \textbf{(a)} held-out internal NIH test set, \textbf{(b)} external CheXpert dataset, and \textbf{(c)} external MIMIC dataset. Comparison between the original dataset and ISLE's optimal decomposition is annotated. For all other comparisons, only statistically significant differences are annotated. (ns: $p \geq 0.05$, *: $p < 0.05$, **: $p < 0.01$, ***: $p < 0.001$)}
    \label{fig:auc_results}
\end{figure*}

On the held-out internal NIH test set, the optimal decomposition was determined as the third decomposition out of the total five with resolution 256x256, which corresponds to the closest resolution to the model's input (i.e., 224x224). The optimal decomposition yielded a comparable AUROC of $0.838 \pm 0.050$ compared to the AUROC of $0.838 \pm 0.052$ measured with the original PNG data ($p=0.62$). ISLE required 91.58\% less data to be transmitted, $87.65\%$ less time to decode, and increased throughput by $375.76\%$. The results are detailed in Fig. \ref{fig:auc_results}(a) and Table \ref{tab:nih_results}.

When evaluated on unseen data from the CheXpert dataset, the optimal decomposition yielded a comparable AUROC of $0.709 \pm 0.082$ compared to the AUROC of $0.699 \pm 0.078$ measured with the original JPEG data ($p=0.99$). ISLE required 97.43\% less data to be transmitted, 97.70\% less time to decode, and increased throughput by 2403.04\%. The results are detailed in Fig. \ref{fig:auc_results}(b) and Table \ref{tab:cxpt_results}.

Similarly, when evaluated on unseen data from the MIMIC dataset, the optimal decomposition yielded a comparable AUROC of $0.771 \pm 0.072$ compared to the AUROC of $0.762 \pm 0.078$ measured with the original JPEG data ($p=0.99$). More importantly, ISLE required 98.62\% less data to be transmitted, 98.49\% less time to decode, and increased throughput by 3056.01\%. When comparing ISLE with the original DICOM version of MIMIC, we observe no significant differences in the AUROC ($0.779 \pm 0.068, p=0.99$) while transmitting 99.83\% less data, taking 99.52\% less time to decode, and increasing throughput by 11,387.17\%. The results are detailed in Fig. \ref{fig:auc_results}(c) and Table \ref{tab:mimic_results}.

\begin{table}[!t]
    \centering
    \caption{Data metrics for the held-out internal NIH test set. We compare the metrics of ISLE's optimal decomposition with the original PNG format. Metrics for HTJ2K without the ISLE's stream optimization are also included.}
    \label{tab:nih_results}
    \small
    \begin{tabular*}{\linewidth}{@{\extracolsep{\fill}} lccc}
          & \textbf{Data Transferred} & \textbf{Decode Time} & \textbf{Throughput} \\  
          & \textbf{(GB)} & \textbf{(s)} & \textbf{(images/s)} \\ \toprule
        PNG & 8.60 & 121.72 & 165.59 \\
        \midrule
        HTJ2K & 8.26 (\textbf{-4\%}) & 120.50 (\textbf{-1\%}) & 169.14 (\textbf{+2\%}) \\
        ISLE & 0.72 (\textbf{-92\%}) & 15.03 (\textbf{-88\%}) & 787.78 (\textbf{+376\%}) \\
        \bottomrule
    \end{tabular*}
\end{table}

\begin{table}[!t]
    \centering
    \caption{Data metrics for the external CheXpert dataset. We compare the metrics of ISLE's optimal decomposition with the original JPEG format. Metrics for HTJ2K without the ISLE's stream optimization are also included.}
    \label{tab:cxpt_results}
    \small
    \begin{tabular*}{\linewidth}{@{\extracolsep{\fill}} lccc}
          & \textbf{Data Transferred} & \textbf{Decode Time} & \textbf{Throughput} \\  
          & \textbf{(GB)} & \textbf{(s)} & \textbf{(images/s)} \\ \toprule
        JPEG & 31.59 & 535.69 & 27.05 \\
        \midrule
        HTJ2K & 52.08 (\textbf{+65\%}) & 402.42 (\textbf{-25\%}) & 36.26 (\textbf{+34\%}) \\
        ISLE & 0.81 (\textbf{-97\%}) & 12.34 (\textbf{-98\%}) & 676.99 (\textbf{+2403\%}) \\
        \bottomrule
    \end{tabular*}
\end{table}

\begin{table}[!t]
    \centering
    \caption{Data metrics for the external MIMIC dataset. We compare the metrics of ISLE's optimal decomposition with the original JPEG format. Metrics for the DICOM version of MIMIC and HTJ2K without the ISLE's stream optimization are also included.}
    \label{tab:mimic_results}
    \small
    \begin{tabular*}{\linewidth}{@{\extracolsep{\fill}} lccc}
          & \textbf{Data Transferred} & \textbf{Decode Time} & \textbf{Throughput} \\  
          & \textbf{(GB)} & \textbf{(s)} & \textbf{(images/s)} \\ \toprule
        JPEG & 24.78 & 491.92 & 29.43 \\
        DCM & 199.24 & 1540.88 & 8.09 \\
        \midrule
        HTJ2K & 43.94 (\textbf{+77\%}) & 438.59 (\textbf{-11\%}) & 33.42 (\textbf{+14\%}) \\
        ISLE & 0.34 (\textbf{-99\%}) & 7.44 (\textbf{-98\%}) & 928.76 (\textbf{+3056\%}) \\
        \bottomrule
    \end{tabular*}
\end{table}

\section{Discussion}

Our results demonstrate the excellent potential of ISLE in dramatically reducing the data transmission, compute, and time required to decode imaging data for medical imaging AI inference. ISLE significantly increased throughput for medical imaging AI inference with no significant impact on the AI system's performance. One of the primary advantages of ISLE is the ability to stream images to different AI systems at different resolutions without any additional image processing or computational requirements. This would allow healthcare systems to store their imaging data as a single copy and at a single resolution while streaming that data to different AI systems and providers at different resolutions. 

As the adoption of AI systems within the clinical environment grows, smaller practices and rural clinics may not be able to keep pace with larger hospital systems in terms of implementing high-quality computational infrastructure and therefore may not be able to support deployment of large-scale AI algorithms leading to widening of health disparities in the US population \cite{douthit2015exposing}. The proposed ISLE framework could potentially help maintain equity in care provided by different healthcare providers across the country by reducing the computational infrastructural requirement for healthcare providers by more than 98\%.  

Our work has certain limitations. In this work, we developed and evaluated ISLE for 2D CXR classification. The current framework is not capable of compressing and streaming 3D medical imaging data (e.g., MRI, CT). In addition, ISLE has not yet been evaluated for other AI tasks such as segmentation and object detection. In the future, we plan to further develop the ISLE framework to deal with 3D medical imaging data for different AI applications. In conclusion, ISLE presents an important first step towards high-throughput, compute- and bandwidth- optimized, cost effective AI inference for clinical decision making at scale. 

\section{Compliance with ethical standards} \label{sec:ethics}



This research study was conducted retrospectively using human subject data made available in either open access or credentialed access. The NIH Chest X-Ray 14 dataset is in public domain (\url{https://nihcc.app.box.com/v/ChestXray-NIHCC/folder/36938765345}). The CheXpert dataset has credentialed access under Stanford University Dataset Research Use Agreement (\url{https://stanfordmlgroup.github.io/competitions/chexpert}). The MIMIC-CXR and MIMIC-CXR-JPG datasets have credentialed access under PhysioNet Credentialed Health Data License 1.5.0 (\url{https://physionet.org/content/mimic-cxr-jpg/2.0.0}). Ethical approval is not required as confirmed by the license attached with the datasets.

\section{Acknowledgments} \label{sec:acknowledgments}




No funding was received for conducting this study. Authors P.K., A.K., E.S., P.H.Y., and V.S.P. hold a US provisional patent in the intelligent streaming technology described in this work.

\bibliographystyle{IEEEbib}
\bibliography{refs}

\end{document}